\documentclass[preprint,12pt]{elsarticle}
\usepackage{graphicx}
\usepackage{amssymb}
\begin{document}
\begin{frontmatter}

\author{R. Peng$^{a,b,c}$ and C. B. Yang$^{a,b}$}
\address{$^{a}$Institute of Particle Physics, Hua-Zhong Normal
University, Wuhan 430079, People's Republic of China\\
$^{b}$ Key Laboratory of Quark and Lepton (Hua-Zhong Normal University),
Ministry of Education, People's Republic of China\\
$^c$College of Science, Wuhan University of Science and Technology, Wuhan 430065, People's Republic of China\\
Corresponding author. E-mail address: pengru$\_$1204@hotmail.com}

\title{$\pi$-$J/\psi$ Correlation and Elliptic Flow Parameter $v_2$ of Charmed Mesons at RHIC Energy}

\begin{abstract}
We study the correlation between the trigger $\pi$ and the associated $J/\psi$ on near and away sides in Au+Au collisions at $\sqrt{s_{NN}}=200$ GeV. In the region of trigger momentum $p_t>4$ GeV/$c$, the $\pi$ spectrum is composed of thermal-shower and shower-shower recombinations in the frame work of the recombination model. We consider the azimuthal anisotropy in the quenched hard parton distribution and then calculate the elliptic flow parameter $v_2$ of charmed mesons ($J/\psi$, $D^0$ and $D_s$) for different centralities.
\end{abstract}
\begin{keyword}
Recombination model; di-hadron correlation; elliptic flow parameter; charmed meson\\
PACS: 25.75.Dw, 25.75.Gz, 25.75.Ld
\end{keyword}
\end{frontmatter}
\section{Introduction}
Jet quenching is one of the most informative probes for the formation
of quark gluon plasma (QGP) in relativistic heavy ion collisions (RHIC). It describes that
partons produced in the initial collisions with high $p_T$ lose a large fraction of their energies when they propagate through the matter before the formation of final state hadrons. The experimental results of the suppression of both inclusive hadron \cite{PHENIX1,STAR1,PHENIX2}
and correlated away-side di-hadron yields \cite{STAR2} were predicted from this energy loss
\cite{jetquen1,jetquen2}. Usually the energy loss of a jet is caused by two main mechanisms:
elastic collisions with deconfined partons and induced gluon radiation. Since the energy loss is related to the traverse path length of the jet \cite{length1,length2}, two-particle correlation is effective in investigating jet-medium interaction corresponding to the passage of a hard or semihard parton \cite{correlation1}. In Ref.\cite{energyloss} the correlation of $\pi$-$\pi$ in jets produced on near and away sides of high $p_T$ triggers has been studied in the frame work of the recombination model. Now we consider the heavy-flavored meson yields and calculate the di-hadron correlation of $\pi$-$J/\psi$ in this paper.

The elliptic flow which is a measure of the anisotropic momentum distribution of the produced particles is another important observable at RHIC. The elliptical flow ($v_2$) results from the initial geometrical anisotropy which creates pressure gradients soon after the start of the hydrodynamic evolution. Thus, the final $v_2$ is sensitive to the fireball evolution and allows us to probe the dynamical properties of the dense matter produced in the collisions. Recently PHENIX Collaboration has measured the $J/\psi$ elliptic flow parameter $v_2$ in Au+Au collisions at $\sqrt{s_{NN}}=200$ GeV \cite{PHENIXv2}.
In this paper, we calculate $v_2$ in the recombination model and compare our results with the experimental data.

This paper is organized as follows. We calculate the di-hadron correlation of $\pi$-$J/\psi$ and $v_2$ of charmed mesons in Section 2 and 3, respectively. And a brief summary is given in the final section.

\section{Di-hadron correlation of $\pi$-$J/\psi$}
Based on the recombination model \cite{energyloss} we have calculated the transverse momentum spectrum of $J/\psi$ which fits the experimental data well \cite{pengru1}, and also have predicted $p_T$ spectra of other heavy flavored mesons for different centralities \cite{pengru2}. In this section we apply the shower parton distributions (SPD) to calculate the di-hadron correlation of $\pi$-$J/\psi$ on near and away sides. We use $p$ as a generic symbol for the transverse momentum of any hadron, $p_t$ for the trigger particle ($\pi$), $p_a$ for the associated particle ($J/\psi$) on the near side and $p_b$ for the associated particle ($J/\psi$) on the away side.

In the recombination model the meson production is expressed as the sum of three terms, $\mathcal{TT}$ (pure thermal), $\mathcal{TS}$ (thermal-shower) and $\mathcal{SS}$ (shower-shower), since there are two components of parton sources: thermal ($\mathcal{T}$) partons and shower ($\mathcal{S}$) partons which originate from hard partons. Thus we have
\begin{equation}
\label{production}
\frac{dN_{M}}{d^2p}=\frac{d(N_{M}^{\mathcal{TT}}+N_{M}^{\mathcal{TS}}
+N_{M}^{\mathcal{SS}})}{d^2p}.
\end{equation}

The shower-shower recombination term ($\mathcal{SS}$ contribution) is equivalent to the fragmentation function (FF) \cite{energyloss}
\begin{equation}
\label{SS}
\frac{dN^{\mathcal{SS}}_{M}}{pdp}=\frac{1}{p^{0}p}\sum_{i}\int\frac{dq}{q}F_{i}(q)
\frac{p}{q}D^{M}_{i}(\frac{p}{q}),
\end{equation}
where
\begin{equation}
\label{hardparton1}
F_{i}(q)=\frac{1}{\beta L}\int^{qe^{\beta L}}_{q}dkkf_{i}(k)
\end{equation}
is the distribution of hard parton $i$ in Au+Au collisions at $\sqrt{s_{NN}}=200$ GeV after traversing the medium of length $L$. The distribution $f_{i}(k)=dN^{hard}_{i}/d^2kdy$ of hard parton $i$ just after hard scattering at mid-rapidity can be found in Refs.\cite{FGLUON} and \cite{FCHARM}. $\beta L$ is the explicit dynamical medium factor to describe the energy loss effect and its value for different centralities is determined by fitting the single-pion inclusive distribution \cite{energyloss}.
$D^{M}_{i}$ is the FF of parton $i$ splitting into meson $M$ and the results of FFs for $\pi$ and $J/\psi$ are obtained in Ref.\cite{AKK} and Refs.\cite{EBTCY,TCY}, respectively.

The thermal parton transverse momentum distribution density is given in Ref.\cite{RBC}
\begin{equation}
\mathcal{T}(p)=\frac{dN^{th}}{dp^{2}dy}\bigg|_{y=0}\bigg/(\tau A_T ).
\end{equation}
It's assumed that hadronization occurs at $\tau=5$ fm with temperature $T=175$ MeV in the parton phase \cite{RBC}, which is consistent with predictions of the phase transition temperature at vanishing baryon chemical potential from lattice QCD \cite{QCD}. $A_{T}=\rho^{2}_0\pi$ is the transverse area of the parton system with the radius $\rho_0=9$ fm \cite{RBC}. Using the parameterized SPD $S^{j}_{i}(z)$ \cite{pengru1,SPD}, we can determine the distribution of shower parton $j$ with transverse momentum $q_1$ in central Au+Au collisions as \cite{energyloss}
\begin{equation}
\label{shower}
\mathcal{S}(q_1)=\sum_{i}\int\frac{dq}{q}F'_{i}(q)S^{j}_{i}(q_1/q),
\end{equation}
where
\begin{equation}
F'_{i}(q)=\frac{1}{\beta L}\int^{qe^{\beta L}}_{q}\frac{dk}{k}f_{i}(k)/E_i
\end{equation}
is a little different from Eq.(\ref{hardparton1}) \cite{pengru1}. Then the $\mathcal{TS}$ term can be calculated as \cite{pengru2}
\begin{equation}
\label{TS1}
\frac{dN^{\mathcal{TS}}}{pdp}=\sum_{i}\int\frac{dq}{q}F_{i}(q)
\widehat{TS}(q,p),
\end{equation}
where
\begin{equation}
\label{TS2}
\widehat{TS}(q,p)=C_{M}(2\pi)\int^{1}_{0}dx\frac{\arrowvert\phi_{M}(x)\arrowvert^2}{k^2E_i}
[\frac{\mathcal{T}_{a}(q_1)S^{b}_i(q_2/q)}{g\gamma_{a}x}
+\frac{S^{a}_i(q_1/q)\mathcal{T}_{b}(q_2)}
{g\gamma_{b}(1-x)}]
\end{equation}
with the momenta of the two constituent quarks $q_1=xp$ and $q_2=(1-x)p$. $\phi_{M}(x)$ in Eq.(\ref{TS2}) is the wave function of the meson in the momentum space. The wave functions for $\pi$ and $J/\psi$ are determined in Refs.\cite{SPD} and \cite{pengru1}, respectively. $\gamma_a$ and $\gamma_{b}$ stand for the fugacities of the constituent quarks $a$ and $b$. The fugacities of light quarks are $\gamma_u=\gamma_d=1$,  $\gamma_s=0.8$ \cite{RBC} and for charm quark $\gamma_c=0.26$ which is obtained by fitting the $J/\psi$ transverse momentum spectrum \cite{pengru1}. Since every quark has 3 color and 2 spin degrees of freedom, we use the meson degeneracy factor $C_{M}=(3\times 2)^2$.

We consider the di-hadron correlation at intermediate trigger momentum ($p_t>4$ GeV), thus approximately the $\mathcal{TT}$ contribution to the trigger and jet can be neglected. Then the correlation of the particle $J/\psi$ with $p_a$ associated with the trigger particle $\pi$ with $p_t$ on the near side is
\begin{eqnarray}
\label{neardist}
\frac{dN^{\mathrm{near}}_{\pi,J/\psi}}{p_tdp_tp_adp_a}&=&\sum_{i}\int\frac{dq}{q}F_{i}(q)
\{[\widehat{TS}^{\pi}(q,p_t)+\frac{1}{p^0_tq}D^{\pi}_{i}(\frac{p_t}{q})]
\widehat{TS}^{J/\psi}(q-p_t,p_a)\nonumber\\
&  &+\widehat{TS}^{\pi}(q-p_a,p_t)\frac{1}{p^0_aq}D^{J/\psi}_{i}(\frac{p_a}{q})
+\frac{1}{p^0_tp^0_aq^2}D_2(\frac{p_t}{q},\frac{p_a}{q})\},
\end{eqnarray}
where the di-hadron FF $D_2(z_1,z_2)$ is assumed as \cite{energyloss}
\begin{equation}
D_2(z_1,z_2)=\frac{1}{2}[D^{\pi}(z_1)D^{J/\psi}
(\frac{z_2}{1-z_1})+D^{\pi}(\frac{z_1}{1-z_2})D^{J/\psi}(z_2)].
\end{equation}
The di-hadron distribution of the trigger $\pi$ and the associated $J/\psi$ with $p_b$ on the away side is calculated in Ref.\cite{energyloss} as
\begin{eqnarray}
\label{awaydist}
\frac{dN^{\mathrm{away}}_{\pi,J/\psi}}{p_tdp_tp_bdp_b}&=&\frac{e^{\beta L}}{2\beta L}\sum_i\int_{p_t}dq
\int_{q'_0}^{qe^{\beta L}}dq'f_i(\sqrt{qq'e^{\beta L}})
[\widehat{TS}^{\pi}(q,p_t)+\frac{1}{p^0_tq}D^{\pi}_{i}(\frac{p_t}{q})]\nonumber\\
& &\times[\widehat{TS}^{J/\psi}(q',p_b)+\frac{1}{p^0_bq'}D^{J/\psi}_{i}(\frac{p_b}{q'})]
\end{eqnarray}
with $q'_0=\mathrm{Max}(qe^{-\beta L},p_b)$. Here, $q'$ is the momentum of the recoil parton in the trigger jet.

Then the near-side (away-side) yield per trigger for the trigger momentum in a narrow range $\Delta p_t$ ($\Delta p_t\rightarrow0$) around $p_t$ is
\begin{equation}
Y^{\mathrm{near(away)}}_{\pi,J/\psi}=\int_{\Delta p_t}dp_t\frac{dN^{\mathrm{near(away)}}_{\pi,J/\psi}}{p_a(p_b)dp_tdp_a(p_b)}\bigg{/}\int_{\Delta p_t}
dp_t\frac{dN_{\pi}}{dp_t}
\end{equation}
where $dN_{\pi}/dp_t$ is the trigger pion distribution that excludes the $\mathcal{TT}$ component of the inclusive distribution.

\begin{figure}[h]
   \centering
\includegraphics[width=\textwidth]{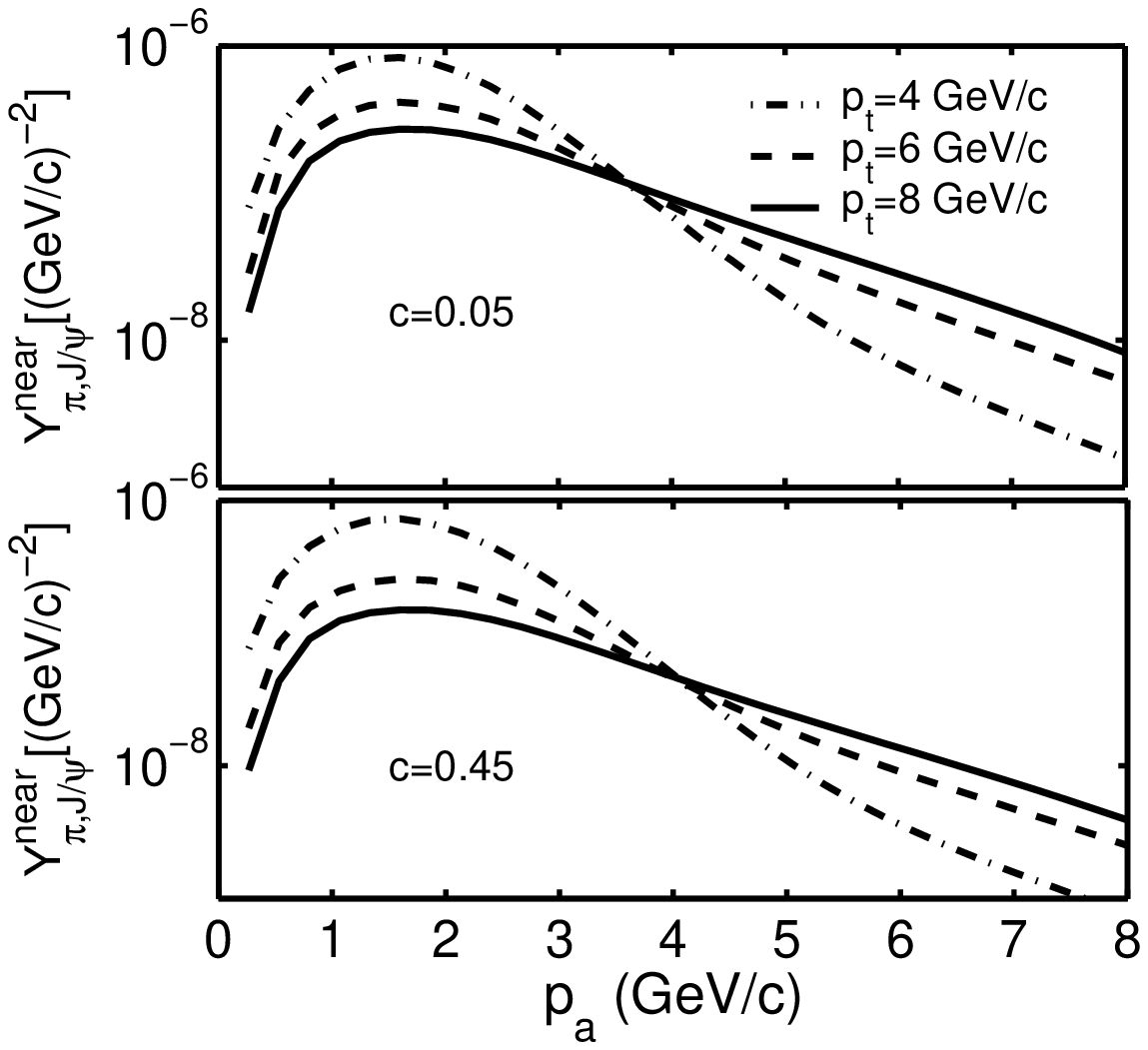}
\caption{Distribution of associated $J/\psi$ ($p_a$) on the near side of a jet triggered by a pion with three momenta ($p_t$) for $c=0.05$ and $c=0.45$.}
\label{near}
\end{figure}

\begin{figure}[h]
   \centering
    \includegraphics[width=\textwidth]{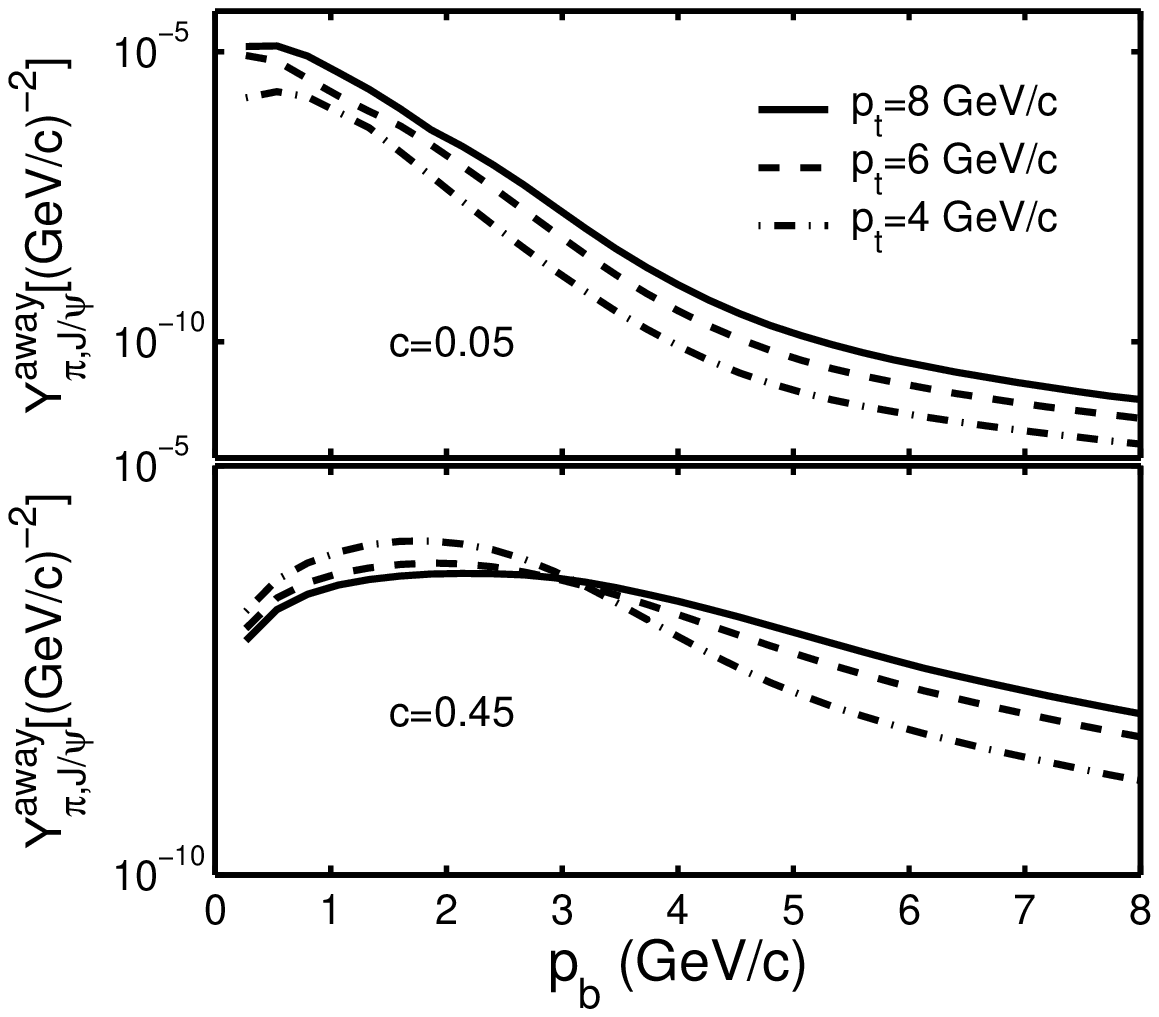}
    \caption{Distribution of associated $J/\psi$ ($p_b$) on the away side of a jet triggered by a pion with three momenta ($p_t$) for $c=0.05$ and $c=0.45$.}
  \label{away}
\end{figure}

We use $c$ to denote the centrality, for example, $c=0.05$ stands for 0-10$\%$ centrality. The  energy loss factor $\beta L$ is dependent on $c$ and its values can be found in Ref.\cite{energyloss}. With the known parameters of $\beta L$ we can calculate the yields for different centralities. The results of the yield of $J/\psi$ on near and away sides per trigger with three values of trigger $\pi$ transverse momentum $p_t$ in Au+Au collisions for 0-10$\%$ and 40-50$\%$ centralities at $\sqrt{s_{NN}}=200$ GeV/$c$ are shown in Fig.\ref{near} and Fig.\ref{away}, respectively. For the central collisions ($c=0.05$), the near-side yield increases with $p_t$ for $p_a>3.8$ GeV/$c$, while this trend is reversed in the region of low $p_a$. The away-side yield becomes higher as $p_t$ increases. We find that the shape of the $p_a$ distribution does not change with the centrality, while the trend of the away-side yield becomes smoother and similar to that of the near-side in more peripheral collisions.

\begin{figure}[h]
   \centering
    \includegraphics[width=\textwidth]{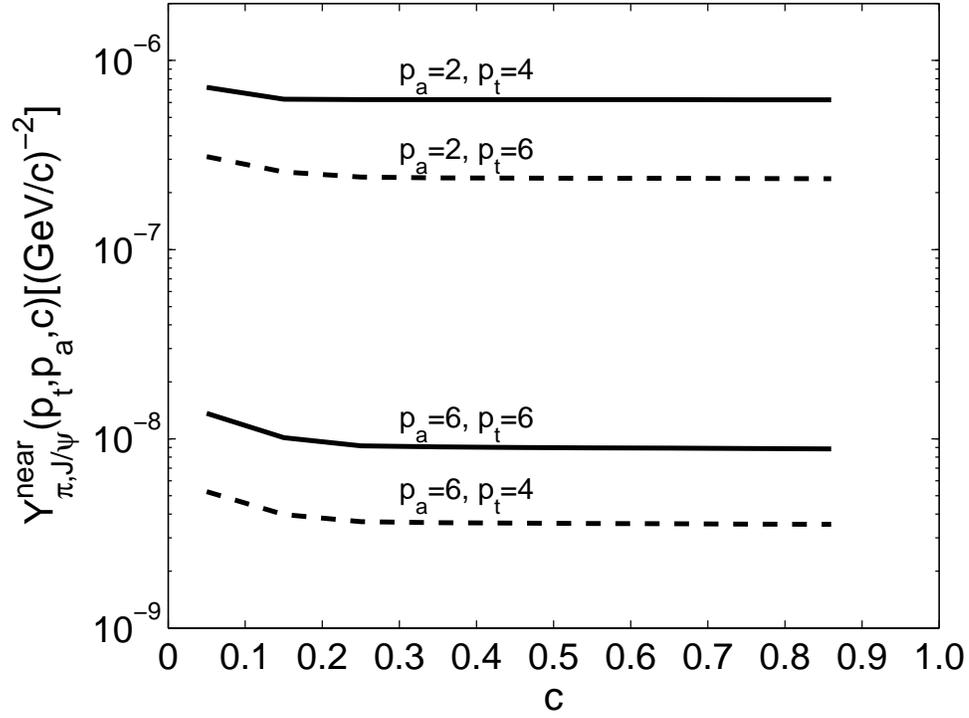}
    \caption{Yield per trigger in the near-side jet as functions of centrality $c$ with $p_a$ and $p_t$ in GeV/$c$.}
  \label{nearL}
\end{figure}

\begin{figure}[h]
   \centering
    \includegraphics[width=\textwidth]{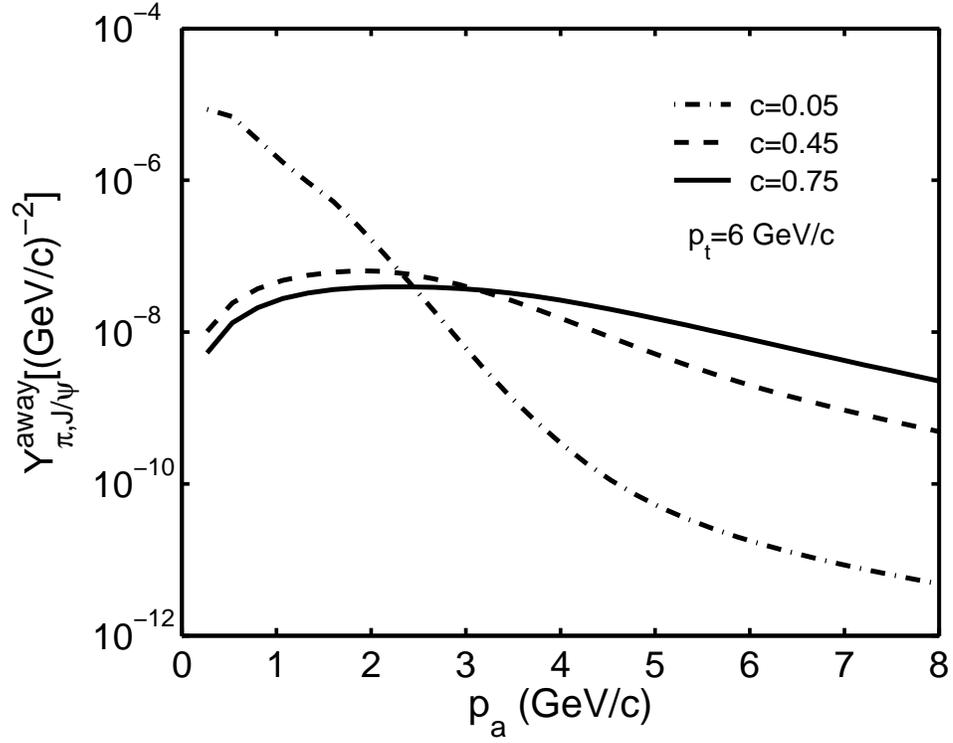}
    \caption{Distribution of associated $J/\psi$ ($p_b$) on the away side of a jet triggered by a pion for three centralities at $p_t=6$ GeV/$c$.}
  \label{awayL}
\end{figure}

In order to clarify the dependence of the yield on the centrality, the near-side yield versus $c$ is shown in Fig.\ref{nearL} for $p_t=4$ and 6 GeV/$c$ and $p_a=2$ and 6 GeV/$c$. The result is nearly a constant in $c$, which has been testified in Ref.\cite{energyloss}. The calculation of the average distance $\langle t\rangle$ the parton travels to reach the near-side surface \cite{energyloss} suggests that the hard scattering point is in a layer roughly 13$\%$ of the medium size $L$ inside the surface and is insensitive to $L$. On average 15$\%$ of the parton energy is lost to the medium, which is also independent on $L$. Thus the centrality dependence of the near-side yield is negligible. The away-side yield increases when $c$ is raised from 0.05 to 0.75 for $p_b>3$ GeV/$c$ as shown in Fig.\ref{awayL}. It has been explained in Ref.\cite{energyloss} when a particle on the away side is required, the scattering point cannot to be too far from the surface of the away side and the point is pulled closer to the surface with higher $c$. So when the nuclear overlap is smaller, it is easier for the recoil jet to reach the away side to produce a particle. The calculation of the ratio of recoil parton average momentum $\langle q'\rangle/\langle k'\rangle$ in Ref.\cite{energyloss} also suggests that the energy loss of the recoil parton becomes smaller with increasing $c$. That means in higher $c$ collisions the recoil parton emerges at the away-side surface with a larger momentum $q'$ at given initial momentum $k'$. Then, at low $p_b$ this leads to a visible smaller momentum fraction $x=p_b/q'$ in Eq.(\ref{awaydist}). The FFs shown in Ref.\cite{SPD} reflect that the probability of $J/\psi$ produced at low $x$ is very small, which results in the lower yield accordingly. Thus the behavior of the dependence on the centrality is quite different in the region of low $p_b$.

The difference between $\pi$-$J/\psi$ and $\pi$-$\pi$ correlation calculated in Ref.\cite{energyloss} is caused by the significantly different FFs for $J/\psi$ and $\pi$. One can take a clear comparison of the FFs in Refs.\cite{pengru1} and \cite{SPD}. It's necessary to note that in Ref.\cite{energyloss} the shower partons can be initiated by hard partons $i=u, d, s, \overline{u}, \overline{d}, \overline{s}$ and gluon. Thus in $\pi$-$\pi$ calculation the FFs for pion come from three terms $D_{v}^{\pi}$, $D_{s}^{\pi}$ and $D_{g}^{\pi}$ where $v$ denote the valence quark, $s$ the sea quark and $g$ the gluon. In $\pi$-$J/\psi$ correlation most contributions are from terms in which $i$ can be $c$ and $g$ since the FFs of light quarks splitting into $J/\psi$ are assumed to be zero in Ref.\cite{EBTCY}. The FF $D^{J/\psi}_c$ increases up to the momentum fraction $x\sim0.75$ ($x\sim0.25$ for $D^{J/\psi}_g$) and then decreases, while FFs for $\pi$ decrease monotonically with $x$. It means that the probability of a parton splitting into $\pi$ is much higher at low $x$ but for $J/\psi$ the larger probability takes place at intermediate $x$. This contrary trend of FFs at low $x$ results in the different behavior of the near-side yield between $\pi$-$\pi$ and $\pi$-$J/\psi$ for $p_a<3.8$ GeV/$c$. This difference also affects the away-side yield at low $p_b$ when $c$ increases with smaller $\beta L$, which is shown in the lower panel of Fig.\ref{away}. The yield for $\pi$-$J/\psi$ is $4\sim5$ orders of magnitude lower than that of $\pi$-$\pi$ since the FFs for $J/\psi$ are much smaller than those for $\pi$ and the hard parton distribution of $c$ is about 3 orders lower than those of light quarks.

What we have calculated is the transverse momentum spectra $dN/pdp$ of $\pi$ and $J/\psi$ averaged over all azimuthal angle $\phi$. Because of the average over $\phi$ in the spectra, only the effective traversing length $\beta L$ appears in the calculations for the medium effect. In the next section, we consider the azimuthal anisotropy in the spectra of charmed mesons and discuss the elliptic flow parameter.

\section{$v_2$ of charmed mesons}
We discuss the azimuthal anisotropy of charmed meson momentum distribution by considering the elliptic flow $v_2$. We assume that the distribution of thermal parton is isotropic so that the recombination of $\mathcal{TT}$ is independent of azimuthal angle ($\phi$). Then the $\phi$ dependence comes from the degraded hard parton distribution, which is caused by the dependence of the traversing length in the medium. Both $\mathcal{TS}$ and $\mathcal{SS}$ components are proportional to the distribution of the hard partons. Thus we can rewrite the meson production (Eq.(\ref{production})) as
\begin{equation}
\label{phi}
\frac{dN_M}{pdpd\phi}(c)=\sum_{i}\int\frac{dq}{q}F_{i}(q,\phi,c)[\widehat{TS}(q,p)+\frac{1}{p^0q}D^M_i
(\frac{p}{q})]+\frac{dN^{\mathcal{TT}}}{pdp},
\end{equation}
where $F_{i}(q,\phi,c)$ is the probability of a hard parton $i$ with momentum $q$ at azimuthal angle $\phi$ in Au+Au collisions with centrality $c$ at $\sqrt{s_{NN}}=200$ GeV.

For a hard parton created with the distribution $f_i(k)$ at the creation point, the initial momentum $k$ changes into $q$ after traversing an absorptive distance $\xi$. The corresponding distribution is expressed by the momentum degradation factor $G(k,q,\xi)$
\begin{equation}
F_i(q,\xi)=\int dkkf_i(k)G(k,q,\xi).
\end{equation}
In terms of $\xi$, $G(k,q,\xi)$ can be written as a simple exponential form $G(k,q,\xi)=q\delta(q-ke^{-\xi})$ \cite{Yang2010}. If $P(\xi,\phi,c)$ is the probability of having a dynamical path length $\xi$ for a parton directed at $\phi$, the hard parton distribution is obtained after carrying out the integration over $\xi$
\begin{equation}
F_{i}(q,\phi,c)=\int d\xi P(\xi,\phi,c)F_{i}(q,\xi).
\end{equation}
In Ref.\cite{Yang2010}, a scaling behavior of $P(\xi,\phi,c)$ is found for the dependencies on $\phi$ and $c$ for pion production. $P(\xi,\phi,c)$ can be written as a universal function in terms of a scaling variable $z=\xi/\overline{\xi}(\phi,c)$
\begin{equation}
P(\xi,\phi,c)=\psi(z)/\overline{\xi}(\phi,c),
\end{equation}
where $\overline{\xi}$ is the mean dynamical length. The scaling function is parameterized by
\begin{equation}
\psi(z)=\zeta^{a_1}(1-\zeta)^{a_2}/B(a_1+1,a_2+1), \zeta=z/2.4
\end{equation}
with $a_1=0.37$ and $a_2=0.81$, where $B(\alpha,\beta)$ is the Beta function. We can get the results of the average dynamical path length $\overline{\xi}$ as a function of $\phi$ and $c$ in Ref.\cite{Yang2010}.

The elliptic flow is quantified by the second Fourier coefficient $v_2$ of the meson azimuthal distribution
\begin{equation}
\frac{dN}{p_Tdp_Td\phi}=A(p_T)[1+2v_2(p_T)\cos(2\phi)].
\end{equation}
Then we can get the elliptic flow parameter $v_2$ of charmed mesons ($J/\psi$, $D^0$, $D_s$) dependent on transverse momentum for different centralities.

\begin{figure}[h]
\centering
\includegraphics[width=\textwidth]{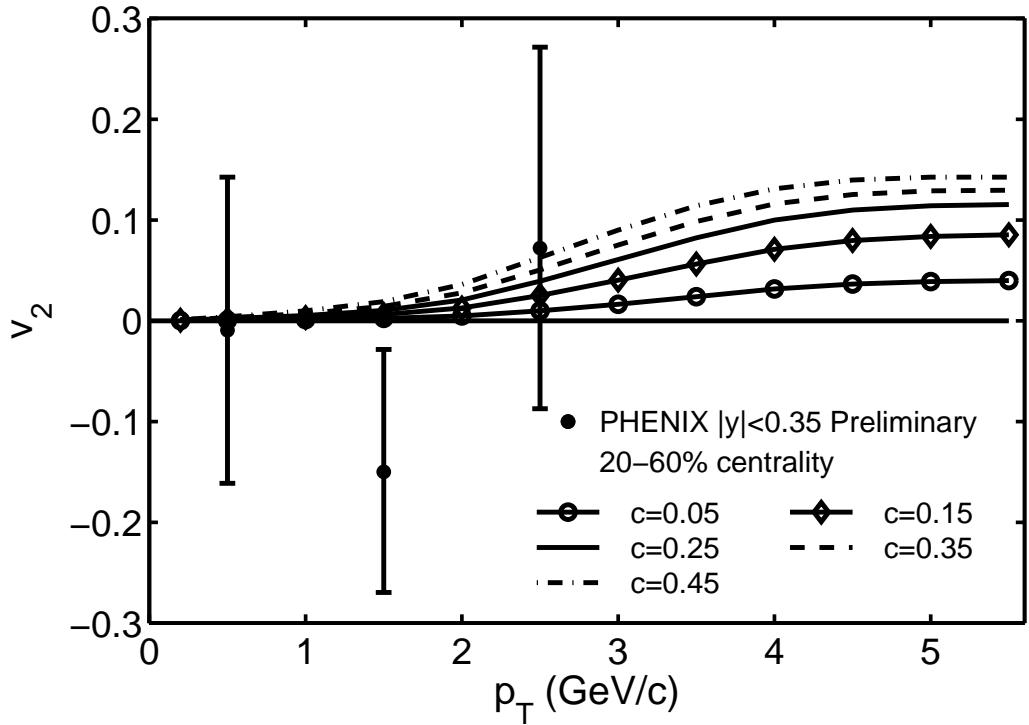}
\caption{$v_2$ for $J/\psi$ as a function of $p_T$ in Au+Au collisions at $\sqrt{s_{NN}}=200$ GeV for different centralities. The data shown for three $p_T$'s are from \cite{PHENIXv2}.}
\label{Jpsiv2}
\end{figure}

The results of $J/\psi$ $v_2$ for different centralities are exhibited in Fig.\ref{Jpsiv2} with the experimental data from PHENIX Collaboration \cite{PHENIXv2}. The calculated results are in agreement with the data within errors except the negative point at $p_T=1.5$ GeV/$c$. $v_2$ increases monotonically up to $p_T\simeq 4$ GeV/$c$ and then starts to saturate to constants  $0.04-0.14$ from $c=0.05-0.45$, depending on the centrality. And the values of $v_2$ get larger from $0.05-0.45$ centralities at given $p_T$. The trend of $p_T$ and centrality dependencies of $v_2$ is similar to that of other charge hadrons shown by PHENIX in Ref.\cite{PHENIXv2light}.

\begin{figure}[h]
   \centering
    \includegraphics[width=\textwidth]{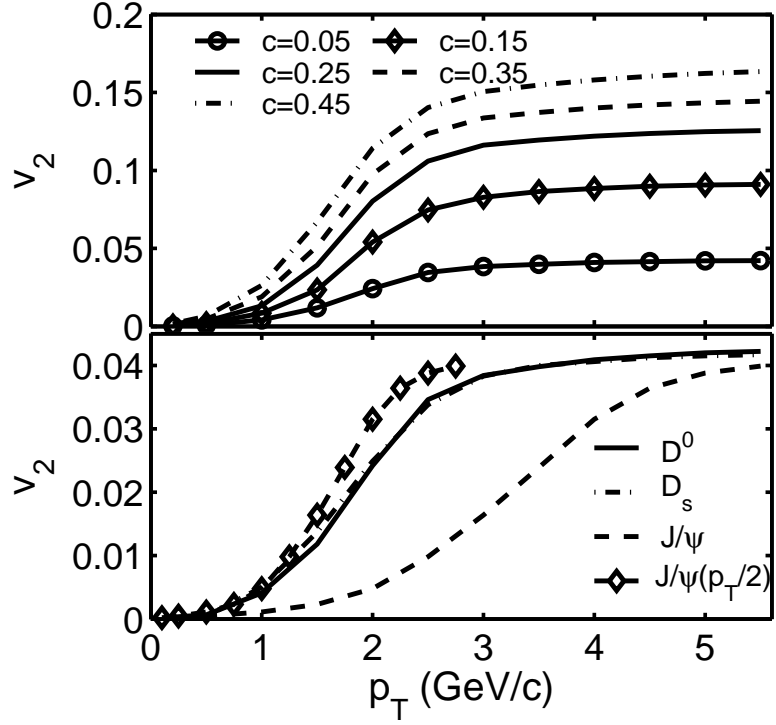}
    \caption{$v_2$ for $D^0$ as a function of $p_T$ in Au+Au collisions at $\sqrt{s_{NN}}=200$ GeV for different centralities and the comparison of the elliptic flow parameter $v_2$ for three charmed mesons (c=0.05).}
  \label{comparev2}
\end{figure}

We also predict $v_2$ of $D^0$ and $D_s$ and the results for $D^0$ are shown in Fig.\ref{comparev2} and the comparison of $v_2$ for the three different charmed mesons at $c=0.05$ is exhibited in the lower panel. Obviously, $v_2$ of $D^0$ or $D_s$ is much larger than that of $J/\psi$. This feature results from the different constituent quark mass in the three mesons. The dependence of momentum distribution on $\phi$ in Eq.(\ref{phi}) comes from two terms $\mathcal{TS}$ and $\mathcal{SS}$ with the hard parton distribution $F_i(q,\phi,c)$. The thermal parton distribution of light quarks ($u$, $d$ and $s$) is much higher than that of heavy quark \cite{pengru1}. Thus, for $D^0$ the main contribution to $\mathcal{TS}$ comes from $\mathcal{T}_{\overline{u}}\mathcal{S}_c$. And its transverse momentum is mainly from that of the charm quark. Roughly one can take $p_{D^0}\simeq p_{c}$. For $J/\psi$, since its two constituent quarks have the same mass, the transverse momentum for both $c$ and $\overline{c}$ in $J/\psi$ are the same as half of that of $J/\psi$. Thus one would expect that $v^{J/\psi}_2(p_T/2)\simeq v^{D^0}_2(p_T)$. And $v^{J/\psi}_2(p_T/2)$ is shown in the lower panel of Fig.\ref{comparev2} by the dashed line with symbols. $v_2$ for $D^0$ and $D_s$ is almost the same and larger than that of $J/\psi$ at the same $p_T$.

The number of quark (NQ) scaling of $v_2$, $v_2/n$ versus $p_T/n$, has been observed at RHIC \cite{v2scalingSTAR,v2scalingPHENIX}. The scaling lends strong support to the finding that the collectivity develops in the partonic stage at RHIC. Quark coalescence model has been used to explain the flavor dependence of $v_2$ \cite{v2scalingmodel}. In this model hadrons are combined by the coalescing quarks and then $v_2$ of hadrons can be expressed in terms of $v_2$ of quarks, where there are no difference in collectivity among $u$, $d$ and $s$ quarks. Such a scaling implies that the momentum of the hadron is simply the sum of the momentum of the coalescing quarks. In our model the momentum of $D^0$ or $D_s$ comes mainly form charm quark due to the unequal constituent quark masses $m_c\gg m_u(m_s)$. Thus we get different $v_2$ for $J/\psi$ and $D^0$ or $D_s$. Our prediction can be tested by further experiments. If the experimental data get the larger value of $v_2$ for $D^0$ as we have expected, it offers a good proof for the success of the recombination model. So the violation of the $v_2$ NQ scaling also cannot prove the non-formation of QGP in the collisions.

\section{Conclusions}
The correlation of $\pi$-$J/\psi$ and the elliptic flow parameter $v_2$ of charmed mesons are studied in the frame work of the recombination model. In the region of trigger particle momentum $p_t>4$ GeV/$c$, the contribution to the meson production comes from $\mathcal{TS}$ and $\mathcal{TS}$ recombination ($\mathcal{TT}$ contribution can be neglected). The away-side yield per trigger increases with the trigger momentum $p_t$. And the difference between $\pi$-$\pi$ and $\pi$-$J/\psi$ correlation takes place in the near-side yield in the region of $p_a<3.8$ GeV/$c$ and the away-side yield in collisions with higher centrality at low $p_b$, which is caused by the different behavior of the fragmentation functions for $\pi$ and $J/\psi$.

The elliptic flow parameter $v_2$ of charmed mesons for different centralities is obtained by considering the azimuthal anisotropy in the hard parton distribution $F_i(q,\phi,c)$. $v_2$ increases up to $p_t\sim 3.5-4$ GeV/$c$ and then saturates. $v_2$ of $D^0$ and $D_s$ is almost the same in the whole $p_T$ region and larger than that of $J/\psi$.

\newpage

\newpage

\newpage

\newpage

\newpage

\newpage

\end{document}